\documentclass[aps,prd,twocolumn,superscriptaddress,nofootinbib]{revtex4-1}

\usepackage{latexsym}
\usepackage{amsmath}
\usepackage{amssymb}
\usepackage{amsfonts}
\usepackage{bm}

\usepackage{color}
\definecolor{purple}{rgb}{0.5,0,0.5}
\definecolor{blue}{rgb}{0.0,0,0.9}
\definecolor{prdblue}{rgb}{0.133,0.118,0.498}
\usepackage[colorlinks=true, pdfstartview=FitV, linkcolor=prdblue, citecolor= prdblue, urlcolor=prdblue]{hyperref}

\usepackage{supertabular} 
\usepackage{placeins}
\usepackage{epsfig}
\usepackage{graphicx}

\begin{document}

%\preprint{}

\title{The $\mathbf{X(3872)}$'s excitation and its connection with production at hadron colliders}


\author{M.C. Gordillo}
\email[]{cgorbar@upo.es}
\affiliation{Departamento de Sistemas F\'isicos, Qu\'imicos y Naturales, Universidad Pablo de Olavide, E-41013 Sevilla, Spain}

\author{F. De Soto}
\email[]{fcsotbor@upo.es}
\affiliation{Departamento de Sistemas F\'isicos, Qu\'imicos y Naturales, Universidad Pablo de Olavide, E-41013 Sevilla, Spain}

\author{J. Segovia}
\email[]{jsegovia@upo.es}
\affiliation{Departamento de Sistemas F\'isicos, Qu\'imicos y Naturales, Universidad Pablo de Olavide, E-41013 Sevilla, Spain}


\date{\today}

\begin{abstract}
The LHCb collaboration has found that the production rate of $X(3872)$ in proton-proton collisions decreases as final state particle multiplicity increases. Moreover, the ALICE experiment at CERN has observed that the number of deuterons produced increases with multiplicity, a behavior that is qualitatively different from that of the $X(3872)$. These experimental findings may point to a compact structure for the $X(3872)$ or, at least, that its hadronization could proceed through a charm-anticharm core.
We have recently used a diffusion Monte Carlo method to solve the many-body Schr\"odinger equation that describes the $X(3872)$ as a $c \bar c q \bar q$ tetraquark system with quantum numbers $I^G(J^{PC})=0^+(1^{++})$ and $1^-(1^{++})$. According to our structural analysis, the quark--(anti-)quark correlations resemble light-meson--heavy-meson molecules of type $\omega J/\psi$ and $\rho J/\psi$, rather than the most extended $D\bar D^{\ast}$ interpretation. It was argued that this fact may be the key to make compatible the molecular features of the $X(3872)$ with its production observables.
The same formalism allows us to compute the first color excited $c \bar c q \bar q$ tetraquark state with either $I^G(J^{PC})=0^+(1^{++})$ or $1^-(1^{++})$. A bound-state is found in each channel, their masses are around 4.0 GeV which is an energy region where many new exotic candidates have been collected by the Particle Data Group. Concerning their structural properties, these states cluster in a compact diquark-antidiquark arrangement which matches perfectly with a so-called Born-Oppenheimer tetraquark configuration. The promptly production rates of these states in proton-proton, proton-nucleus and nucleus-nucleus collisions should fall off equal to or even faster than those of the $X(3872)$.
\end{abstract}

%\pacs{
%12.38.-t \and % Quantum Chromodynamics
%12.39.-x \and % Potential Models

\maketitle

%%%%%%%%%%%%%%%%%%%%%%%%%%%%%%%%%%%%%%%%%%%%%%%%%%%%%%%%%%%%%%%%%%%%%%%%%%%%%%%%
%%%%%%%%%%%%%%%%%%%%%%%%%%%%%%%%%%%%%%%%%%%%%%%%%%%%%%%%%%%%%%%%%%%%%%%%%%%%%%%%

\noindent\emph{Introduction}.\,---\, 
The structure of exotic states, called collectively XYZ, has been a matter of intense scientific debate during the last two decades~\cite{Brambilla:2010cs, Esposito:2016noz, Chen:2016qju, Lebed:2016hpi, Guo:2017jvc, Olsen:2017bmm, Brambilla:2019esw}. The $X(3872)$ is the most studied among them, observed in 2003 by the Belle collaboration as an unexpected peak in the $\pi^+ \pi^- J/\psi$ invariant mass spectrum of the decay $B^+ \to K^+ \pi^+ \pi^- J/\psi$~\cite{Belle:2003nnu}. Its mass is almost exactly at the $D^0\bar{D}^{\ast0}$ threshold and it is remarkably narrow~\cite{ParticleDataGroup:2020ssz}. The pion pair is dominated by the $\rho$-meson, thus showing sizable isospin violation, unexpected if the $X(3872)$ were a coventional charmonium state.

Since the simple charm-anticharm structure cannot account for the observed features of $X(3872)$, more valence quarks are needed. Its minimal content would be $c\bar c q\bar q$, with $q$ either $u$- or $d$-quark, and the additional quarks could be gather together by color forces forming a new kind of hadron, which is basically classified into a compact tetraquark of hadronic size~\cite{Maiani:2004vq, Brodsky:2014xia} or a hadron molecule with an extension larger than $1\,\text{fm}$~\cite{Braaten:2003he, Close:2003sg, Tornqvist:2004qy}. Concerning the second case, it is advocated that the $X(3872)$ is a loosely-bound $DD^\ast$-molecule, the meson counterpart of the deuteron, due to its closeness with respect to threshold. Other molecular arrangements for the $X(3872)$ have not been frequently considered, despite the existence of alternatives such as hadro-charmonium~\cite{Dubynskiy:2008mq}, which consists of a compact color-singlet $(c\bar{c})$-pair surrounded by a color-singlet $(q\bar{q})$-pair bounded through color (van der Waals) interactions; see also the proposal of Ref.~\cite{Braaten:2013boa} related with the so-called Born-Oppenheimer tetraquarks that consist on a color-octet $(c\bar{c})$-pair and a color-octet $(q\bar{q})$-pair bounded by the exchange of gluons.

The LHCb collaboration has recently observed that the production rate of promptly produced $X(3872)$, relative to the $\psi(2S)$, as a function of final state particle multiplicity, decreases with increasing multiplicity~\cite{LHCb:2020sey}. An effect that is firmly known to affect the production of ordinary heavy quarkonia in proton-nucleus collisions due to final state breakup interactions between quarkonia and co-moving particles~\cite{Ferreiro:2014bia, Ferreiro:2018wbd}. Moreover, the ALICE collaboration has recently published an analysis for deuteron production in proton-proton collisions~\cite{ALICE:2019dgz, ALICE:2020foi} showing that the number of deuterons produced increases with multiplicity, a behavior that is qualitatively different from that of the $X(3872)$. The idea that interactions with co-movers could favor the coalescence of a hadron molecule was originally proposed in~\cite{Esposito:2013ada, Guerrieri:2014gfa} for nucleon-nucleon, and in~\cite{ExHIC:2010gcb, Cho:2013rpa} for nucleus-nucleus collisions.

The $DD^\ast$ molecular interpretation of $X(3872)$ can avoid the mentioned experimental challenges if its hadronization proceeds through a compact $c\bar c$ core, which is difficult to assume when the $X(3872)$ description is based simply on the same nuclear forces that bind together two nucleons to form the deuteron. While maintaining hadron molecular picture, one can resort to different configurations as hadro-charmonium to explain the new observed properties related with $X(3872)$ production.\footnote{Note herein that compact tetraquark interpretations suffer from other deficiencies such as an overpopulation of exotic states.} In Ref.~\cite{Gordillo:2021bra}, we used a diffusion Monte Carlo method to solve the many-body Schr\"odinger equation that describes the $X(3872)$ as a $c \bar c q \bar q$ tetraquark. Two $c\bar c q \bar q$ loosely-lying states with quantum numbers $I^G(J^{PC})=0^+(1^{++})$ and $1^-(1^{++})$ were found. According to our results, the two quarks and two antiquarks are arranged as light-meson--heavy-meson molecules of type $\omega J/\psi$ and $\rho J/\psi$, rather than the most extended $D\bar D^{\ast}$ interpretation. This fact would be the key to make compatible the molecular features of the $X(3872)$ with its decay and production observables that seem to indicate the presence of a $c\bar c$ cluster.

Among other advantages, the diffusion Monte Carlo avoids the usual quark-clustering assumed in any theoretical technique applied to the same problem, which is crucial in the study of $X(3872)$'s nature. Moreover, the interaction between particles was modeled by the most general and accepted potential, \emph{i.e.} a pairwise interaction including Coulomb, linear-confining and hyperfine spin-spin terms. Goldstone-boson exchange interactions between light quarks were also considered. However, they played a marginal role: the chiral contribution to the mass of the $X(3872)$ represented only $10\%$, and also the chiral potentials are so weak that does not produce meson-meson molecular states by themselves.

Our theoretical formalism allows to study (color) excited tetraquark configurations from the lowest-lying states reported in Ref.~\cite{Gordillo:2021bra}. This letter is devoted to present their masses, wave functions and structural properties, highlighting the fact that its nature is completely different from our theoretical candidates of the $X(3872)$ meson.

%%%%%%%%%%%%%%%%%%%%%%%%%%%%%%%%%%%%%%%%%%%%%%%%%%%%%%%%%%%%%%%%%%%%%%%%%%%%%%%%

\noindent\emph{Theoretical formalism}.\,---\,
The use of Quantum Monte Carlo (QMC) methods to hadron physics has been scarce~\cite{Carlson:1982xi, Carlson:1983rw, Bai:2016int} because these tools are ideally suited for many-body physics~\cite{Hammond:1994bk, Foulkes:2001zz, Nightingale:2014bk} and most known hadrons consist only on 2- and 3-body bound states, {\it i.e.} mesons and baryons. The quark model paradigm of hadrons is, however, changing in the last twenty years with many experimental signals pointing out the possible existence of a new particle zoo made of tetra-, penta- and even hexa-quark systems~\cite{ParticleDataGroup:2020ssz}.

We used a Diffusion Monte Carlo (DMC) method to solve the non-relativistic bound-state problem of fully-heavy tetraquark systems in Ref.~\cite{Gordillo:2020sgc}. The dynamics was driven by a 2-body potential consisting on Coulomb$\,+\,$linear-confining$\,+\,$hyperfine spin-spin terms whose parameters were constrained by a simultaneous fit of $36$ mesons and $53$ baryons~\cite{Semay:1994ht, SilvestreBrac:1996bg}. We demonstrated (see left-bottom panel of Fig.~7 in Ref.~\cite{Gordillo:2020sgc}) that the $J^{PC}=1^{++}$ $cb\bar c\bar b$ ground state prefers to be organized in clusters of $c\bar c$ and $b\bar b$, whose extensions are less than $0.5\,\text{fm}$, separated by a distance of about $0.8-1.0\,\text{fm}$. This arrangement of quarks (antiquarks) is not imitated by its $cb\bar c\bar b$ tetraquark partners with different quantum numbers $J^{PC}=0^{++}$, $1^{+-}$, and $2^{++}$; neither seen in any other explored case of fully-heavy tetraquarks.

Motivated by such theoretical observation, the DMC method was soon after applied to the $J^{PC}=1^{++}$ $cq\bar c\bar q$ system, in the isoscalar and isovector sectors~\cite{Gordillo:2021bra}. Unlike fully-heavy tetraquarks, an additional dynamical mechanism must be taken into account: Goldstone-boson exchange potentials~\cite{Fernandez:1993hx, Valcarce:1995dm, Vijande:2004he, Segovia:2008zza}. Their expressions can be found in, {\it e.g.}, Ref.~\cite{Segovia:2011dg} and have been fixed in the last 10-20 years reproducing hadron~\cite{Segovia:2009zz, Segovia:2011zza, Segovia:2015dia}, hadron-hadron~\cite{Ortega:2016pgg, Ortega:2018cnm, Ortega:2021xst} and multiquark phenomenology~\cite{Yang:2018oqd, Yang:2020twg, Yang:2020fou}. Note, too, that the DMC algorithm allows us to use the same formulae without considering the regularization procedure, \emph{i.e.} we take the limit $\Lambda_\chi\to \infty$ in the mentioned expressions.

It is worth mentioning herein that the set of model parameters are fitted to reproduce a certain number of hadron observables within a determinate range of agreement with experiment. Therefore, it is difficult to assign an error to those parameters and, as a consequence, to the magnitudes calculated when using them. As the range of agreement between theory and experiment is around $(10-20)\%$, this value can be taken as an estimation of the model uncertainty.

%%%%%%%%%%%%%%%%%%%%%%%%%%%%%%%%%%%%%%%%%%%%%%%%%%%%%%%%%%%%%%%%%%%%%%%%%%%%%%%%
\begin{figure*}
\includegraphics[width=0.45\textwidth]{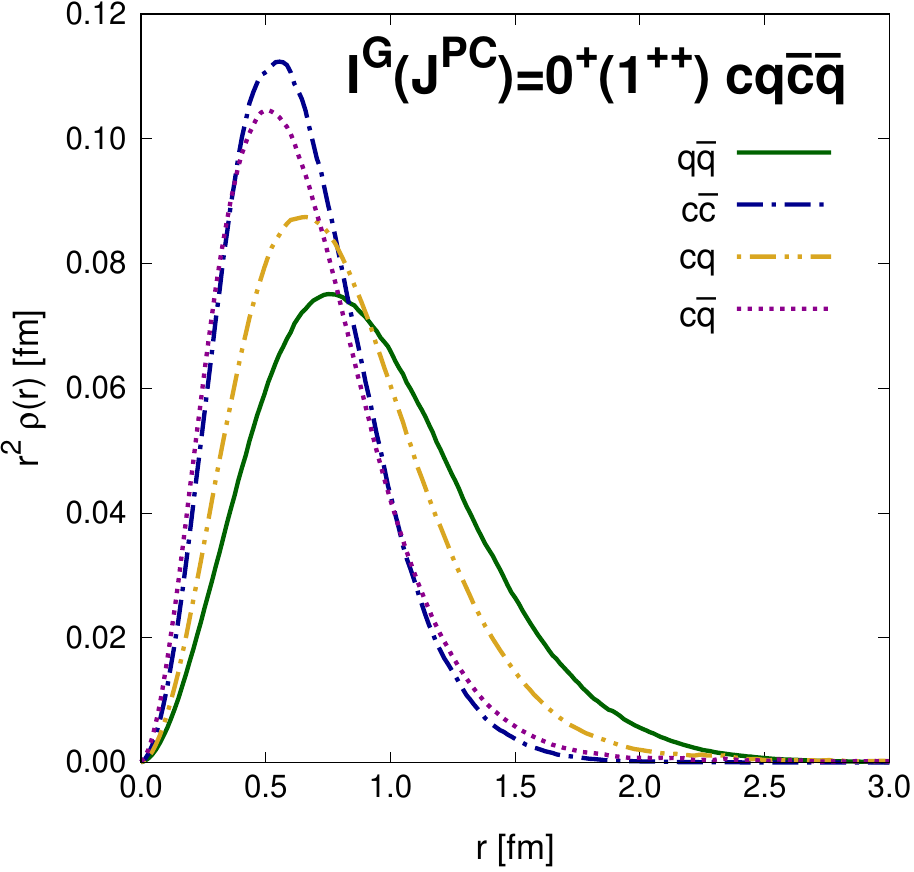}
\includegraphics[width=0.45\textwidth]{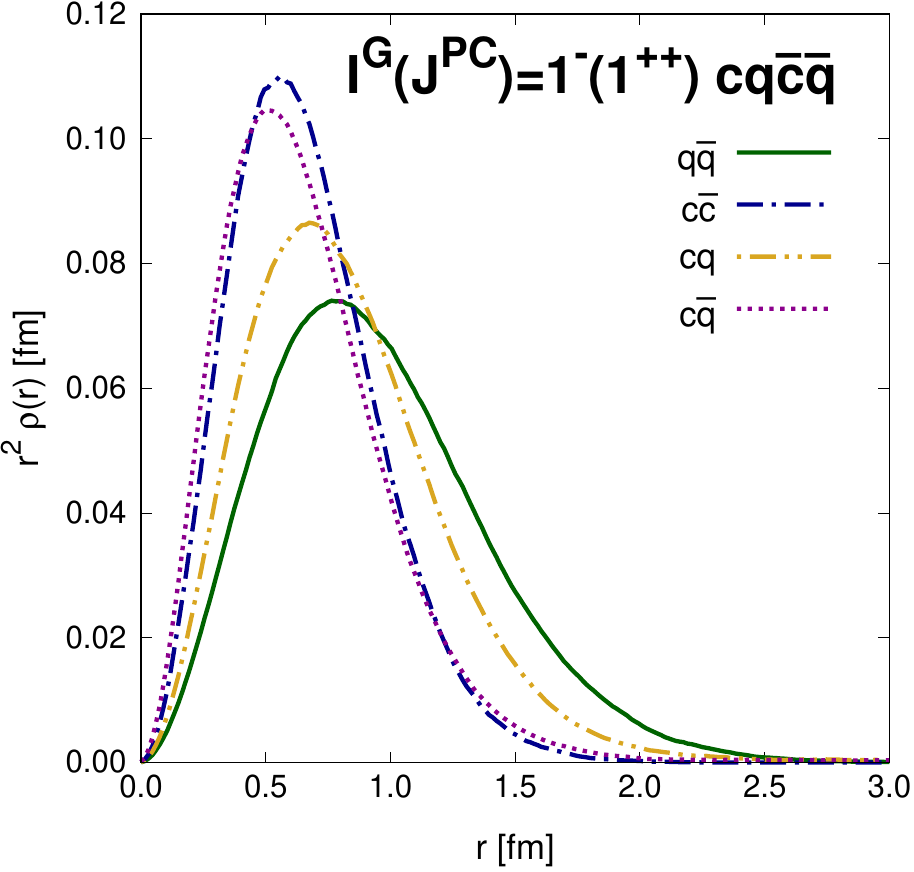}
\caption{\label{fig:Correlations} Radial distribution functions of the first color excited $c \bar c q \bar q$ tetraquark state with either $I^G(J^{PC})=0^+(1^{++})$ (left panel) or $I^G(J^{PC})=1^-(1^{++})$ (right panel) quantum numbers. These functions represent the probability of finding the 2 quarks (antiquarks) at an interquark distance $r$. In both panels, solid (green), dot-dashed (blue), dot-dot-dashed (yellow) and dotted (purple) represent, respectively, $q\bar q$, $c\bar c$, $cq$ and $c\bar q$ correlations inside the $c \bar c q \bar q$ tetraquark.}
\end{figure*}
%%%%%%%%%%%%%%%%%%%%%%%%%%%%%%%%%%%%%%%%%%%%%%%%%%%%%%%%%%%%%%%%%%%%%%%%%%%%%%%%

We obtained two $c\bar c q \bar q$ loosely-lying states with quantum numbers $I^G(J^{PC})=0^+(1^{++})$ and $1^-(1^{++})$, whose masses were, respectively, $3834\,\text{MeV}$ and $3842\,\text{MeV}$. In order to get agreement with the $X(3872)$'s experimental mass, the used quark masses~\cite{SilvestreBrac:1996bg}, $m_q=315\,\text{MeV}$ and $m_c=1836\,\text{MeV}$, can be fine-tuned. In any case, the model uncertainty allows well to assert that theoretical and experimental masses are in fair agreement. These states could contribute separately to the $X(3872)$ signal, or being explained by a coupling between them. As one can see in Fig.~1 of Ref.~\cite{Gordillo:2021bra}, the significant feature is that these states prefer to be arranged as light-meson--heavy-meson molecules of type $\omega J/\psi$ and $\rho J/\psi$, which may explain the observed promptly production properties of the $X(3872)$. In fact, the $X(3872)$'s associated color wave function satisfies:
\begin{align}
|X(3872) \rangle_{\text{color}} = 0.57 \, |\bar{3}_{qc} 3_{\bar q\bar c}\rangle_{\text{color}} + 0.82 \, |6_{qc} \bar{6}_{\bar q \bar c}\rangle_{\text{color}} \,,
\label{eq:X3872wf}
\end{align}
in both isoscalar and isovector $J^{PC}=1^{++}$ channels. Moreover, using Eqs.~(56) and~(57) of Ref.~\cite{Gordillo:2020sgc}, Eq.~\eqref{eq:X3872wf} can be translated into
\begin{align}
|X(3872) \rangle_{\text{color}} &\approx |1_{c\bar c} \bar{1}_{q\bar q} \rangle_{\text{color}} \,,
\end{align}
indicating that the computed tetraquarks prefer to be in a color-singlet $(c\bar{c})$-pair plus a color-singlet $(q\bar{q})$-pair configurations.

\noindent\emph{The $X(3872)$'s first (color) excitation}.\,---\,
The same formalism allows us to compute the first color excitation of the $X(3872)$, interpreted as $c \bar c q \bar q$ tetraquark state. The most simple way to proceed is fixing an orthogonal color wave function:
\begin{align}
|X^\prime \rangle_{\text{color}} = - 0.82 \, |\bar{3}_{qc} 3_{\bar q\bar c}\rangle_{\text{color}} + 0.57 \, |6_{qc} \bar{6}_{\bar q \bar c}\rangle_{\text{color}} \,,
\end{align}
which, using again Eqs.~(56) and~(57) of Ref.~\cite{Gordillo:2020sgc}, can be translated to
\begin{align}
|X^\prime \rangle_{\text{color}} &\approx | 8_{c\bar c} \bar{8}_{q\bar q}  \rangle_{\text{color}} \,.
\label{eq:Xpwf}
\end{align}
It basically assumes a structure named quarkonium adjoint meson or Born-Oppenheimer tetraquark~\cite{Braaten:2013boa} for the first excitation of the $X(3872)$. This consists of two bounded color-octets, $(c\bar{c})_{8}$ and $(q\bar{q})_{8}$, with dynamics similar to 
that of quark-gluon hybrids~\cite{Meyer:2015eta}.

We obtain for the first color excited state of the $J^{PC}=1^{++}$ $c \bar c q \bar q$ system the binding energy $-302\,\text{MeV}$, in the isoscalar sector, and $-294\,\text{MeV}$, in the isovector one. Note also that such values correspond to the absolute masses $4000\,\text{MeV}$ and $4008\,\text{MeV}$. The PDG~\cite{ParticleDataGroup:2020ssz} reports the observation of 9 states in the mass range $[3.9,4.1]\,\text{GeV}$, which makes the {\it naive} quark model picture unable to encompass all of them. Some of these states are well established experimentally while the nature of others is completely unknown, without even being clear about their quantum numbers. Therefore, one could assign to our theoretical states any of the experimental signals as, for instance, the $X(4050)^\pm$ candidate; but strong statements related with any assignment would not be suitable at this time.

We turn now our attention on the structure of above bound-states, exploiting the concept of radial distribution function because it provides valuable information of the existence of interquark correlations; in particular, 2-body correlations. Figure~\ref{fig:Correlations} shows the radial distribution functions of the first color excited $c\bar c q \bar q$ tetraquark state with either $I^G(J^{PC})=0^+(1^{++})$ (left panel) or $I^G(J^{PC})=1^-(1^{++})$ (right panel). These functions represent the probability of finding two quarks (antiquarks) at an interquark distance $r$. In both panels, solid (green), dot-dashed (blue), dot-dot-dashed (yellow) and dotted (purple) represent, respectively, $q\bar q$, $c\bar c$, $cq$ and $c\bar q$ correlations inside the $c \bar c q \bar q$ tetraquark. All radial distribution functions have a mean value less than 1 fm indicating that quark--(anti-)quark correlations are mainly short distance effects of a compact object.

Looking at our results, one can conclude: (i) the color-excited $J^{PC}=1^{++}$ $c \bar c q \bar q$ tetraquark state tends to cluster in a diquark-antidiquark configuration which resembles a Born-Oppenheimer tetraquark; (ii) every quark--(anti-)quark correlation has an extension $\lesssim1\,\text{fm}$, indicating that it is a compact object; (iii) contrary to the ground state, there is no trivial connection between quark--(anti-)quark radial distributions and any kind of meson wave functions; and (iv) $q\bar q$, $c\bar c$ and $c\bar q$ and $c q$ correlations fall off to zero with the interquark distance, indicating the fact of having finite size for the calculated tetraquark $c\bar cq\bar q$ structures.

Our theoretical interpretation of the $X(3872)$ as a hadro-charmonium consisting on a compact color-singlet $(c\bar c)$-pair surrounded by a color-singlet $(q\bar q)$-pair bounded through color interactions could be tested looking for the abundance of its excited (color) state in high-multiplicity pp collisions. Since $X^\prime$ is interpreted as a Born-Oppenheimer $J^{PC}=1^{++}$ $c \bar c q \bar q$ tetraquark state at around 4.0 GeV, one should expect that its production yield decreases with respect to final state particle multiplicity, even in a larger rate than in the case of $X(3872)$ because the quarks and antiquarks are closer. The LHCb and ALICE experiments at CERN are in the position of performing such kind of investigations.

%%%%%%%%%%%%%%%%%%%%%%%%%%%%%%%%%%%%%%%%%%%%%%%%%%%%%%%%%%%%%%%%%%%%%%%%%%%%%%%%

\noindent\emph{Epilogue}.\,---\, We use a diffusion Monte Carlo method to solve the many-body Schr\"odinger equation that describes the color excited $J^{PC}=1^{++}$ $c \bar c q \bar q$ tetraquark system, in both isoscalar and isovector sectors. Among other advantages, this approach avoids the usual quark-clustering assumed in any theoretical technique applied to the same problem and, moreover, provides information about the hadron's wave function and structural properties.

Two bound-states were found whose masses at around 4.0 GeV make them perfect candidates for any of the states collected by the PDG in such energy region. Concerning their structural properties, these states cluster in a diquark-antidiquark configuration which resembles a Born-Oppenheimer tetraquark. Any quark--(anti-)quark correlation has an extension $\lesssim1\,\text{fm}$, pointing out to be compact objects. Moreover, such correlations fall off to zero indicating the fact of having finite size for the calculated tetraquark $c\bar cq\bar q$ structures. Finally, contrary to the ground states investigated in Ref.~\cite{Gordillo:2021bra} and assigned to the $X(3872)$ signal, quark--(anti-)quark radial distributions do not follow any kind of meson wave functions.

The production rates of these states in proton-proton, proton-nucleus and nucleus-nucleus collisions as a function of final state particle
multiplicity could, first, confirm our interpretation of the $X(3872)$ as a hadro-charmonium and, second, the compact nature of the excitations with production rates that should fall off equal to or even faster than those of the $X(3872)$. Let us stress again that such kind of experiments can be already performed by the LHCb and ALICE collaborations at CERN.

%%%%%%%%%%%%%%%%%%%%%%%%%%%%%%%%%%%%%%%%%%%%%%%%%%%%%%%%%%%%%%%%%%%%%%%%%%%%%%%%

\noindent\emph{Acknowledgements}.\,---\, This work has been partially funded by the Ministerio Espa\~nol de Ciencia e Innovaci\'on under grant No. PID2019-107844GB-C22; the Junta de Andaluc\'ia under contract No. Operativo FEDER Andaluc\'ia 2014-2020 UHU-1264517 and P18-FR-5057; but also PAIDI FQM-205 and -370. The authors acknowledges, too, the use of the computer facilities of C3UPO at the Universidad Pablo de Olavide, de Sevilla.

%%%%%%%%%%%%%%%%%%%%%%%%%%%%%%%%%%%%%%%%%%%%%%%%%%%%%%%%%%%%%%%%%%%%%%%%%%%%%%%%

\bibliography{DMC-ExcitedX3872}

\begin{thebibliography}{50}%
\makeatletter
\providecommand \@ifxundefined [1]{%
 \@ifx{#1\undefined}
}%
\providecommand \@ifnum [1]{%
 \ifnum #1\expandafter \@firstoftwo
 \else \expandafter \@secondoftwo
 \fi
}%
\providecommand \@ifx [1]{%
 \ifx #1\expandafter \@firstoftwo
 \else \expandafter \@secondoftwo
 \fi
}%
\providecommand \natexlab [1]{#1}%
\providecommand \enquote  [1]{``#1''}%
\providecommand \bibnamefont  [1]{#1}%
\providecommand \bibfnamefont [1]{#1}%
\providecommand \citenamefont [1]{#1}%
\providecommand \href@noop [0]{\@secondoftwo}%
\providecommand \href [0]{\begingroup \@sanitize@url \@href}%
\providecommand \@href[1]{\@@startlink{#1}\@@href}%
\providecommand \@@href[1]{\endgroup#1\@@endlink}%
\providecommand \@sanitize@url [0]{\catcode `\\12\catcode `\$12\catcode
  `\&12\catcode `\#12\catcode `\^12\catcode `\_12\catcode `\%12\relax}%
\providecommand \@@startlink[1]{}%
\providecommand \@@endlink[0]{}%
\providecommand \url  [0]{\begingroup\@sanitize@url \@url }%
\providecommand \@url [1]{\endgroup\@href {#1}{\urlprefix }}%
\providecommand \urlprefix  [0]{URL }%
\providecommand \Eprint [0]{\href }%
\providecommand \doibase [0]{http://dx.doi.org/}%
\providecommand \selectlanguage [0]{\@gobble}%
\providecommand \bibinfo  [0]{\@secondoftwo}%
\providecommand \bibfield  [0]{\@secondoftwo}%
\providecommand \translation [1]{[#1]}%
\providecommand \BibitemOpen [0]{}%
\providecommand \bibitemStop [0]{}%
\providecommand \bibitemNoStop [0]{.\EOS\space}%
\providecommand \EOS [0]{\spacefactor3000\relax}%
\providecommand \BibitemShut  [1]{\csname bibitem#1\endcsname}%
\let\auto@bib@innerbib\@empty
%</preamble>
\bibitem [{\citenamefont {Brambilla}\ \emph {et~al.}(2011)\citenamefont
  {Brambilla} \emph {et~al.}}]{Brambilla:2010cs}%
  \BibitemOpen
  \bibfield  {author} {\bibinfo {author} {\bibfnamefont {N.}~\bibnamefont
  {Brambilla}} \emph {et~al.},\ }\href {\doibase
  10.1140/epjc/s10052-010-1534-9} {\bibfield  {journal} {\bibinfo  {journal}
  {Eur. Phys. J. C}\ }\textbf {\bibinfo {volume} {71}},\ \bibinfo {pages}
  {1534} (\bibinfo {year} {2011})},\ \Eprint {http://arxiv.org/abs/1010.5827}
  {arXiv:1010.5827 [hep-ph]} \BibitemShut {NoStop}%
\bibitem [{\citenamefont {Esposito}\ \emph {et~al.}(2017)\citenamefont
  {Esposito}, \citenamefont {Pilloni},\ and\ \citenamefont
  {Polosa}}]{Esposito:2016noz}%
  \BibitemOpen
  \bibfield  {author} {\bibinfo {author} {\bibfnamefont {A.}~\bibnamefont
  {Esposito}}, \bibinfo {author} {\bibfnamefont {A.}~\bibnamefont {Pilloni}}, \
  and\ \bibinfo {author} {\bibfnamefont {A.~D.}\ \bibnamefont {Polosa}},\
  }\href {\doibase 10.1016/j.physrep.2016.11.002} {\bibfield  {journal}
  {\bibinfo  {journal} {Phys. Rept.}\ }\textbf {\bibinfo {volume} {668}},\
  \bibinfo {pages} {1} (\bibinfo {year} {2017})},\ \Eprint
  {http://arxiv.org/abs/1611.07920} {arXiv:1611.07920 [hep-ph]} \BibitemShut
  {NoStop}%
\bibitem [{\citenamefont {Chen}\ \emph {et~al.}(2016)\citenamefont {Chen},
  \citenamefont {Chen}, \citenamefont {Liu},\ and\ \citenamefont
  {Zhu}}]{Chen:2016qju}%
  \BibitemOpen
  \bibfield  {author} {\bibinfo {author} {\bibfnamefont {H.-X.}\ \bibnamefont
  {Chen}}, \bibinfo {author} {\bibfnamefont {W.}~\bibnamefont {Chen}}, \bibinfo
  {author} {\bibfnamefont {X.}~\bibnamefont {Liu}}, \ and\ \bibinfo {author}
  {\bibfnamefont {S.-L.}\ \bibnamefont {Zhu}},\ }\href {\doibase
  10.1016/j.physrep.2016.05.004} {\bibfield  {journal} {\bibinfo  {journal}
  {Phys. Rept.}\ }\textbf {\bibinfo {volume} {639}},\ \bibinfo {pages} {1}
  (\bibinfo {year} {2016})},\ \Eprint {http://arxiv.org/abs/1601.02092}
  {arXiv:1601.02092 [hep-ph]} \BibitemShut {NoStop}%
\bibitem [{\citenamefont {Lebed}\ \emph {et~al.}(2017)\citenamefont {Lebed},
  \citenamefont {Mitchell},\ and\ \citenamefont {Swanson}}]{Lebed:2016hpi}%
  \BibitemOpen
  \bibfield  {author} {\bibinfo {author} {\bibfnamefont {R.~F.}\ \bibnamefont
  {Lebed}}, \bibinfo {author} {\bibfnamefont {R.~E.}\ \bibnamefont {Mitchell}},
  \ and\ \bibinfo {author} {\bibfnamefont {E.~S.}\ \bibnamefont {Swanson}},\
  }\href {\doibase 10.1016/j.ppnp.2016.11.003} {\bibfield  {journal} {\bibinfo
  {journal} {Prog. Part. Nucl. Phys.}\ }\textbf {\bibinfo {volume} {93}},\
  \bibinfo {pages} {143} (\bibinfo {year} {2017})},\ \Eprint
  {http://arxiv.org/abs/1610.04528} {arXiv:1610.04528 [hep-ph]} \BibitemShut
  {NoStop}%
\bibitem [{\citenamefont {Guo}\ \emph {et~al.}(2018)\citenamefont {Guo},
  \citenamefont {Hanhart}, \citenamefont {Mei\ss{}ner}, \citenamefont {Wang},
  \citenamefont {Zhao},\ and\ \citenamefont {Zou}}]{Guo:2017jvc}%
  \BibitemOpen
  \bibfield  {author} {\bibinfo {author} {\bibfnamefont {F.-K.}\ \bibnamefont
  {Guo}}, \bibinfo {author} {\bibfnamefont {C.}~\bibnamefont {Hanhart}},
  \bibinfo {author} {\bibfnamefont {U.-G.}\ \bibnamefont {Mei\ss{}ner}},
  \bibinfo {author} {\bibfnamefont {Q.}~\bibnamefont {Wang}}, \bibinfo {author}
  {\bibfnamefont {Q.}~\bibnamefont {Zhao}}, \ and\ \bibinfo {author}
  {\bibfnamefont {B.-S.}\ \bibnamefont {Zou}},\ }\href {\doibase
  10.1103/RevModPhys.90.015004} {\bibfield  {journal} {\bibinfo  {journal}
  {Rev. Mod. Phys.}\ }\textbf {\bibinfo {volume} {90}},\ \bibinfo {pages}
  {015004} (\bibinfo {year} {2018})},\ \bibinfo {note} {[Erratum: Rev.Mod.Phys.
  94, 029901 (2022)]},\ \Eprint {http://arxiv.org/abs/1705.00141}
  {arXiv:1705.00141 [hep-ph]} \BibitemShut {NoStop}%
\bibitem [{\citenamefont {Olsen}\ \emph {et~al.}(2018)\citenamefont {Olsen},
  \citenamefont {Skwarnicki},\ and\ \citenamefont {Zieminska}}]{Olsen:2017bmm}%
  \BibitemOpen
  \bibfield  {author} {\bibinfo {author} {\bibfnamefont {S.~L.}\ \bibnamefont
  {Olsen}}, \bibinfo {author} {\bibfnamefont {T.}~\bibnamefont {Skwarnicki}}, \
  and\ \bibinfo {author} {\bibfnamefont {D.}~\bibnamefont {Zieminska}},\ }\href
  {\doibase 10.1103/RevModPhys.90.015003} {\bibfield  {journal} {\bibinfo
  {journal} {Rev. Mod. Phys.}\ }\textbf {\bibinfo {volume} {90}},\ \bibinfo
  {pages} {015003} (\bibinfo {year} {2018})},\ \Eprint
  {http://arxiv.org/abs/1708.04012} {arXiv:1708.04012 [hep-ph]} \BibitemShut
  {NoStop}%
\bibitem [{\citenamefont {Brambilla}\ \emph {et~al.}(2020)\citenamefont
  {Brambilla}, \citenamefont {Eidelman}, \citenamefont {Hanhart}, \citenamefont
  {Nefediev}, \citenamefont {Shen}, \citenamefont {Thomas}, \citenamefont
  {Vairo},\ and\ \citenamefont {Yuan}}]{Brambilla:2019esw}%
  \BibitemOpen
  \bibfield  {author} {\bibinfo {author} {\bibfnamefont {N.}~\bibnamefont
  {Brambilla}}, \bibinfo {author} {\bibfnamefont {S.}~\bibnamefont {Eidelman}},
  \bibinfo {author} {\bibfnamefont {C.}~\bibnamefont {Hanhart}}, \bibinfo
  {author} {\bibfnamefont {A.}~\bibnamefont {Nefediev}}, \bibinfo {author}
  {\bibfnamefont {C.-P.}\ \bibnamefont {Shen}}, \bibinfo {author}
  {\bibfnamefont {C.~E.}\ \bibnamefont {Thomas}}, \bibinfo {author}
  {\bibfnamefont {A.}~\bibnamefont {Vairo}}, \ and\ \bibinfo {author}
  {\bibfnamefont {C.-Z.}\ \bibnamefont {Yuan}},\ }\href {\doibase
  10.1016/j.physrep.2020.05.001} {\bibfield  {journal} {\bibinfo  {journal}
  {Phys. Rept.}\ }\textbf {\bibinfo {volume} {873}},\ \bibinfo {pages} {1}
  (\bibinfo {year} {2020})},\ \Eprint {http://arxiv.org/abs/1907.07583}
  {arXiv:1907.07583 [hep-ex]} \BibitemShut {NoStop}%
\bibitem [{\citenamefont {Choi}\ \emph {et~al.}(2003)\citenamefont {Choi} \emph
  {et~al.}}]{Belle:2003nnu}%
  \BibitemOpen
  \bibfield  {author} {\bibinfo {author} {\bibfnamefont {S.~K.}\ \bibnamefont
  {Choi}} \emph {et~al.} (\bibinfo {collaboration} {Belle}),\ }\href {\doibase
  10.1103/PhysRevLett.91.262001} {\bibfield  {journal} {\bibinfo  {journal}
  {Phys. Rev. Lett.}\ }\textbf {\bibinfo {volume} {91}},\ \bibinfo {pages}
  {262001} (\bibinfo {year} {2003})},\ \Eprint
  {http://arxiv.org/abs/hep-ex/0309032} {arXiv:hep-ex/0309032} \BibitemShut
  {NoStop}%
\bibitem [{\citenamefont {Zyla}\ \emph {et~al.}(2020)\citenamefont {Zyla} \emph
  {et~al.}}]{ParticleDataGroup:2020ssz}%
  \BibitemOpen
  \bibfield  {author} {\bibinfo {author} {\bibfnamefont {P.~A.}\ \bibnamefont
  {Zyla}} \emph {et~al.} (\bibinfo {collaboration} {Particle Data Group}),\
  }\href {\doibase 10.1093/ptep/ptaa104} {\bibfield  {journal} {\bibinfo
  {journal} {PTEP}\ }\textbf {\bibinfo {volume} {2020}},\ \bibinfo {pages}
  {083C01} (\bibinfo {year} {2020})}\BibitemShut {NoStop}%
\bibitem [{\citenamefont {Maiani}\ \emph {et~al.}(2005)\citenamefont {Maiani},
  \citenamefont {Piccinini}, \citenamefont {Polosa},\ and\ \citenamefont
  {Riquer}}]{Maiani:2004vq}%
  \BibitemOpen
  \bibfield  {author} {\bibinfo {author} {\bibfnamefont {L.}~\bibnamefont
  {Maiani}}, \bibinfo {author} {\bibfnamefont {F.}~\bibnamefont {Piccinini}},
  \bibinfo {author} {\bibfnamefont {A.~D.}\ \bibnamefont {Polosa}}, \ and\
  \bibinfo {author} {\bibfnamefont {V.}~\bibnamefont {Riquer}},\ }\href
  {\doibase 10.1103/PhysRevD.71.014028} {\bibfield  {journal} {\bibinfo
  {journal} {Phys. Rev. D}\ }\textbf {\bibinfo {volume} {71}},\ \bibinfo
  {pages} {014028} (\bibinfo {year} {2005})},\ \Eprint
  {http://arxiv.org/abs/hep-ph/0412098} {arXiv:hep-ph/0412098} \BibitemShut
  {NoStop}%
\bibitem [{\citenamefont {Brodsky}\ \emph {et~al.}(2014)\citenamefont
  {Brodsky}, \citenamefont {Hwang},\ and\ \citenamefont
  {Lebed}}]{Brodsky:2014xia}%
  \BibitemOpen
  \bibfield  {author} {\bibinfo {author} {\bibfnamefont {S.~J.}\ \bibnamefont
  {Brodsky}}, \bibinfo {author} {\bibfnamefont {D.~S.}\ \bibnamefont {Hwang}},
  \ and\ \bibinfo {author} {\bibfnamefont {R.~F.}\ \bibnamefont {Lebed}},\
  }\href {\doibase 10.1103/PhysRevLett.113.112001} {\bibfield  {journal}
  {\bibinfo  {journal} {Phys. Rev. Lett.}\ }\textbf {\bibinfo {volume} {113}},\
  \bibinfo {pages} {112001} (\bibinfo {year} {2014})},\ \Eprint
  {http://arxiv.org/abs/1406.7281} {arXiv:1406.7281 [hep-ph]} \BibitemShut
  {NoStop}%
\bibitem [{\citenamefont {Braaten}\ and\ \citenamefont
  {Kusunoki}(2004)}]{Braaten:2003he}%
  \BibitemOpen
  \bibfield  {author} {\bibinfo {author} {\bibfnamefont {E.}~\bibnamefont
  {Braaten}}\ and\ \bibinfo {author} {\bibfnamefont {M.}~\bibnamefont
  {Kusunoki}},\ }\href {\doibase 10.1103/PhysRevD.69.074005} {\bibfield
  {journal} {\bibinfo  {journal} {Phys. Rev. D}\ }\textbf {\bibinfo {volume}
  {69}},\ \bibinfo {pages} {074005} (\bibinfo {year} {2004})},\ \Eprint
  {http://arxiv.org/abs/hep-ph/0311147} {arXiv:hep-ph/0311147} \BibitemShut
  {NoStop}%
\bibitem [{\citenamefont {Close}\ and\ \citenamefont
  {Page}(2004)}]{Close:2003sg}%
  \BibitemOpen
  \bibfield  {author} {\bibinfo {author} {\bibfnamefont {F.~E.}\ \bibnamefont
  {Close}}\ and\ \bibinfo {author} {\bibfnamefont {P.~R.}\ \bibnamefont
  {Page}},\ }\href {\doibase 10.1016/j.physletb.2003.10.032} {\bibfield
  {journal} {\bibinfo  {journal} {Phys. Lett. B}\ }\textbf {\bibinfo {volume}
  {578}},\ \bibinfo {pages} {119} (\bibinfo {year} {2004})},\ \Eprint
  {http://arxiv.org/abs/hep-ph/0309253} {arXiv:hep-ph/0309253} \BibitemShut
  {NoStop}%
\bibitem [{\citenamefont {Tornqvist}(2004)}]{Tornqvist:2004qy}%
  \BibitemOpen
  \bibfield  {author} {\bibinfo {author} {\bibfnamefont {N.~A.}\ \bibnamefont
  {Tornqvist}},\ }\href {\doibase 10.1016/j.physletb.2004.03.077} {\bibfield
  {journal} {\bibinfo  {journal} {Phys. Lett. B}\ }\textbf {\bibinfo {volume}
  {590}},\ \bibinfo {pages} {209} (\bibinfo {year} {2004})},\ \Eprint
  {http://arxiv.org/abs/hep-ph/0402237} {arXiv:hep-ph/0402237} \BibitemShut
  {NoStop}%
\bibitem [{\citenamefont {Dubynskiy}\ and\ \citenamefont
  {Voloshin}(2008)}]{Dubynskiy:2008mq}%
  \BibitemOpen
  \bibfield  {author} {\bibinfo {author} {\bibfnamefont {S.}~\bibnamefont
  {Dubynskiy}}\ and\ \bibinfo {author} {\bibfnamefont {M.~B.}\ \bibnamefont
  {Voloshin}},\ }\href {\doibase 10.1016/j.physletb.2008.07.086} {\bibfield
  {journal} {\bibinfo  {journal} {Phys. Lett. B}\ }\textbf {\bibinfo {volume}
  {666}},\ \bibinfo {pages} {344} (\bibinfo {year} {2008})},\ \Eprint
  {http://arxiv.org/abs/0803.2224} {arXiv:0803.2224 [hep-ph]} \BibitemShut
  {NoStop}%
\bibitem [{\citenamefont {Braaten}(2013)}]{Braaten:2013boa}%
  \BibitemOpen
  \bibfield  {author} {\bibinfo {author} {\bibfnamefont {E.}~\bibnamefont
  {Braaten}},\ }\href {\doibase 10.1103/PhysRevLett.111.162003} {\bibfield
  {journal} {\bibinfo  {journal} {Phys. Rev. Lett.}\ }\textbf {\bibinfo
  {volume} {111}},\ \bibinfo {pages} {162003} (\bibinfo {year} {2013})},\
  \Eprint {http://arxiv.org/abs/1305.6905} {arXiv:1305.6905 [hep-ph]}
  \BibitemShut {NoStop}%
\bibitem [{\citenamefont {Aaij}\ \emph {et~al.}(2021)\citenamefont {Aaij} \emph
  {et~al.}}]{LHCb:2020sey}%
  \BibitemOpen
  \bibfield  {author} {\bibinfo {author} {\bibfnamefont {R.}~\bibnamefont
  {Aaij}} \emph {et~al.} (\bibinfo {collaboration} {LHCb}),\ }\href {\doibase
  10.1103/PhysRevLett.126.092001} {\bibfield  {journal} {\bibinfo  {journal}
  {Phys. Rev. Lett.}\ }\textbf {\bibinfo {volume} {126}},\ \bibinfo {pages}
  {092001} (\bibinfo {year} {2021})},\ \Eprint
  {http://arxiv.org/abs/2009.06619} {arXiv:2009.06619 [hep-ex]} \BibitemShut
  {NoStop}%
\bibitem [{\citenamefont {Ferreiro}(2015)}]{Ferreiro:2014bia}%
  \BibitemOpen
  \bibfield  {author} {\bibinfo {author} {\bibfnamefont {E.~G.}\ \bibnamefont
  {Ferreiro}},\ }\href {\doibase 10.1016/j.physletb.2015.07.066} {\bibfield
  {journal} {\bibinfo  {journal} {Phys. Lett. B}\ }\textbf {\bibinfo {volume}
  {749}},\ \bibinfo {pages} {98} (\bibinfo {year} {2015})},\ \Eprint
  {http://arxiv.org/abs/1411.0549} {arXiv:1411.0549 [hep-ph]} \BibitemShut
  {NoStop}%
\bibitem [{\citenamefont {Ferreiro}\ and\ \citenamefont
  {Lansberg}(2018)}]{Ferreiro:2018wbd}%
  \BibitemOpen
  \bibfield  {author} {\bibinfo {author} {\bibfnamefont {E.~G.}\ \bibnamefont
  {Ferreiro}}\ and\ \bibinfo {author} {\bibfnamefont {J.-P.}\ \bibnamefont
  {Lansberg}},\ }\href {\doibase 10.1007/JHEP10(2018)094} {\bibfield  {journal}
  {\bibinfo  {journal} {JHEP}\ }\textbf {\bibinfo {volume} {10}},\ \bibinfo
  {pages} {094} (\bibinfo {year} {2018})},\ \bibinfo {note} {[Erratum: JHEP 03,
  063 (2019)]},\ \Eprint {http://arxiv.org/abs/1804.04474} {arXiv:1804.04474
  [hep-ph]} \BibitemShut {NoStop}%
\bibitem [{\citenamefont {Acharya}\ \emph {et~al.}(2019)\citenamefont {Acharya}
  \emph {et~al.}}]{ALICE:2019dgz}%
  \BibitemOpen
  \bibfield  {author} {\bibinfo {author} {\bibfnamefont {S.}~\bibnamefont
  {Acharya}} \emph {et~al.} (\bibinfo {collaboration} {ALICE}),\ }\href
  {\doibase 10.1016/j.physletb.2019.05.028} {\bibfield  {journal} {\bibinfo
  {journal} {Phys. Lett. B}\ }\textbf {\bibinfo {volume} {794}},\ \bibinfo
  {pages} {50} (\bibinfo {year} {2019})},\ \Eprint
  {http://arxiv.org/abs/1902.09290} {arXiv:1902.09290 [nucl-ex]} \BibitemShut
  {NoStop}%
\bibitem [{\citenamefont {Acharya}\ \emph {et~al.}(2020)\citenamefont {Acharya}
  \emph {et~al.}}]{ALICE:2020foi}%
  \BibitemOpen
  \bibfield  {author} {\bibinfo {author} {\bibfnamefont {S.}~\bibnamefont
  {Acharya}} \emph {et~al.} (\bibinfo {collaboration} {ALICE}),\ }\href
  {\doibase 10.1140/epjc/s10052-020-8256-4} {\bibfield  {journal} {\bibinfo
  {journal} {Eur. Phys. J. C}\ }\textbf {\bibinfo {volume} {80}},\ \bibinfo
  {pages} {889} (\bibinfo {year} {2020})},\ \Eprint
  {http://arxiv.org/abs/2003.03184} {arXiv:2003.03184 [nucl-ex]} \BibitemShut
  {NoStop}%
\bibitem [{\citenamefont {Esposito}\ \emph {et~al.}(2013)\citenamefont
  {Esposito}, \citenamefont {Piccinini}, \citenamefont {Pilloni},\ and\
  \citenamefont {Polosa}}]{Esposito:2013ada}%
  \BibitemOpen
  \bibfield  {author} {\bibinfo {author} {\bibfnamefont {A.}~\bibnamefont
  {Esposito}}, \bibinfo {author} {\bibfnamefont {F.}~\bibnamefont {Piccinini}},
  \bibinfo {author} {\bibfnamefont {A.}~\bibnamefont {Pilloni}}, \ and\
  \bibinfo {author} {\bibfnamefont {A.~D.}\ \bibnamefont {Polosa}},\ }\href
  {\doibase 10.4236/jmp.2013.412193} {\bibfield  {journal} {\bibinfo  {journal}
  {J. Mod. Phys.}\ }\textbf {\bibinfo {volume} {4}},\ \bibinfo {pages} {1569}
  (\bibinfo {year} {2013})},\ \Eprint {http://arxiv.org/abs/1305.0527}
  {arXiv:1305.0527 [hep-ph]} \BibitemShut {NoStop}%
\bibitem [{\citenamefont {Guerrieri}\ \emph {et~al.}(2014)\citenamefont
  {Guerrieri}, \citenamefont {Piccinini}, \citenamefont {Pilloni},\ and\
  \citenamefont {Polosa}}]{Guerrieri:2014gfa}%
  \BibitemOpen
  \bibfield  {author} {\bibinfo {author} {\bibfnamefont {A.~L.}\ \bibnamefont
  {Guerrieri}}, \bibinfo {author} {\bibfnamefont {F.}~\bibnamefont
  {Piccinini}}, \bibinfo {author} {\bibfnamefont {A.}~\bibnamefont {Pilloni}},
  \ and\ \bibinfo {author} {\bibfnamefont {A.~D.}\ \bibnamefont {Polosa}},\
  }\href {\doibase 10.1103/PhysRevD.90.034003} {\bibfield  {journal} {\bibinfo
  {journal} {Phys. Rev. D}\ }\textbf {\bibinfo {volume} {90}},\ \bibinfo
  {pages} {034003} (\bibinfo {year} {2014})},\ \Eprint
  {http://arxiv.org/abs/1405.7929} {arXiv:1405.7929 [hep-ph]} \BibitemShut
  {NoStop}%
\bibitem [{\citenamefont {Cho}\ \emph {et~al.}(2011)\citenamefont {Cho} \emph
  {et~al.}}]{ExHIC:2010gcb}%
  \BibitemOpen
  \bibfield  {author} {\bibinfo {author} {\bibfnamefont {S.}~\bibnamefont
  {Cho}} \emph {et~al.} (\bibinfo {collaboration} {ExHIC}),\ }\href {\doibase
  10.1103/PhysRevLett.106.212001} {\bibfield  {journal} {\bibinfo  {journal}
  {Phys. Rev. Lett.}\ }\textbf {\bibinfo {volume} {106}},\ \bibinfo {pages}
  {212001} (\bibinfo {year} {2011})},\ \Eprint {http://arxiv.org/abs/1011.0852}
  {arXiv:1011.0852 [nucl-th]} \BibitemShut {NoStop}%
\bibitem [{\citenamefont {Cho}\ and\ \citenamefont {Lee}(2013)}]{Cho:2013rpa}%
  \BibitemOpen
  \bibfield  {author} {\bibinfo {author} {\bibfnamefont {S.}~\bibnamefont
  {Cho}}\ and\ \bibinfo {author} {\bibfnamefont {S.~H.}\ \bibnamefont {Lee}},\
  }\href {\doibase 10.1103/PhysRevC.88.054901} {\bibfield  {journal} {\bibinfo
  {journal} {Phys. Rev. C}\ }\textbf {\bibinfo {volume} {88}},\ \bibinfo
  {pages} {054901} (\bibinfo {year} {2013})},\ \Eprint
  {http://arxiv.org/abs/1302.6381} {arXiv:1302.6381 [nucl-th]} \BibitemShut
  {NoStop}%
\bibitem [{\citenamefont {Gordillo}\ \emph {et~al.}(2021)\citenamefont
  {Gordillo}, \citenamefont {De~Soto},\ and\ \citenamefont
  {Segovia}}]{Gordillo:2021bra}%
  \BibitemOpen
  \bibfield  {author} {\bibinfo {author} {\bibfnamefont {M.~C.}\ \bibnamefont
  {Gordillo}}, \bibinfo {author} {\bibfnamefont {F.}~\bibnamefont {De~Soto}}, \
  and\ \bibinfo {author} {\bibfnamefont {J.}~\bibnamefont {Segovia}},\ }\href
  {\doibase 10.1103/PhysRevD.104.054036} {\bibfield  {journal} {\bibinfo
  {journal} {Phys. Rev. D}\ }\textbf {\bibinfo {volume} {104}},\ \bibinfo
  {pages} {054036} (\bibinfo {year} {2021})},\ \Eprint
  {http://arxiv.org/abs/2105.11976} {arXiv:2105.11976 [hep-ph]} \BibitemShut
  {NoStop}%
\bibitem [{\citenamefont {Carlson}\ \emph
  {et~al.}(1983{\natexlab{a}})\citenamefont {Carlson}, \citenamefont {Kogut},\
  and\ \citenamefont {Pandharipande}}]{Carlson:1982xi}%
  \BibitemOpen
  \bibfield  {author} {\bibinfo {author} {\bibfnamefont {J.}~\bibnamefont
  {Carlson}}, \bibinfo {author} {\bibfnamefont {J.~B.}\ \bibnamefont {Kogut}},
  \ and\ \bibinfo {author} {\bibfnamefont {V.}~\bibnamefont {Pandharipande}},\
  }\href {\doibase 10.1103/PhysRevD.27.233} {\bibfield  {journal} {\bibinfo
  {journal} {Phys. Rev. D}\ }\textbf {\bibinfo {volume} {27}},\ \bibinfo
  {pages} {233} (\bibinfo {year} {1983}{\natexlab{a}})}\BibitemShut {NoStop}%
\bibitem [{\citenamefont {Carlson}\ \emph
  {et~al.}(1983{\natexlab{b}})\citenamefont {Carlson}, \citenamefont {Kogut},\
  and\ \citenamefont {Pandharipande}}]{Carlson:1983rw}%
  \BibitemOpen
  \bibfield  {author} {\bibinfo {author} {\bibfnamefont {J.}~\bibnamefont
  {Carlson}}, \bibinfo {author} {\bibfnamefont {J.}~\bibnamefont {Kogut}}, \
  and\ \bibinfo {author} {\bibfnamefont {V.}~\bibnamefont {Pandharipande}},\
  }\href {\doibase 10.1103/PhysRevD.28.2807} {\bibfield  {journal} {\bibinfo
  {journal} {Phys. Rev. D}\ }\textbf {\bibinfo {volume} {28}},\ \bibinfo
  {pages} {2807} (\bibinfo {year} {1983}{\natexlab{b}})}\BibitemShut {NoStop}%
\bibitem [{\citenamefont {Bai}\ \emph {et~al.}(2019)\citenamefont {Bai},
  \citenamefont {Lu},\ and\ \citenamefont {Osborne}}]{Bai:2016int}%
  \BibitemOpen
  \bibfield  {author} {\bibinfo {author} {\bibfnamefont {Y.}~\bibnamefont
  {Bai}}, \bibinfo {author} {\bibfnamefont {S.}~\bibnamefont {Lu}}, \ and\
  \bibinfo {author} {\bibfnamefont {J.}~\bibnamefont {Osborne}},\ }\href
  {\doibase 10.1016/j.physletb.2019.134930} {\bibfield  {journal} {\bibinfo
  {journal} {Phys. Lett. B}\ }\textbf {\bibinfo {volume} {798}},\ \bibinfo
  {pages} {134930} (\bibinfo {year} {2019})},\ \Eprint
  {http://arxiv.org/abs/1612.00012} {arXiv:1612.00012 [hep-ph]} \BibitemShut
  {NoStop}%
\bibitem [{\citenamefont {Hammond}\ \emph {et~al.}(1994)\citenamefont
  {Hammond}, \citenamefont {Lester},\ and\ \citenamefont
  {Reynolds}}]{Hammond:1994bk}%
  \BibitemOpen
  \bibfield  {author} {\bibinfo {author} {\bibfnamefont {B.}~\bibnamefont
  {Hammond}}, \bibinfo {author} {\bibfnamefont {W.}~\bibnamefont {Lester}}, \
  and\ \bibinfo {author} {\bibfnamefont {P.}~\bibnamefont {Reynolds}},\
  }\href@noop {} {\emph {\bibinfo {title} {Monte Carlo Methods in ab Initio
  Quantum Chemistry}}}\ (\bibinfo  {publisher} {World Scientific},\ \bibinfo
  {address} {Singapore},\ \bibinfo {year} {1994})\BibitemShut {NoStop}%
\bibitem [{\citenamefont {Foulkes}\ \emph {et~al.}(2001)\citenamefont
  {Foulkes}, \citenamefont {Mitas}, \citenamefont {Needs},\ and\ \citenamefont
  {Rajagopal}}]{Foulkes:2001zz}%
  \BibitemOpen
  \bibfield  {author} {\bibinfo {author} {\bibfnamefont {W.}~\bibnamefont
  {Foulkes}}, \bibinfo {author} {\bibfnamefont {L.}~\bibnamefont {Mitas}},
  \bibinfo {author} {\bibfnamefont {R.}~\bibnamefont {Needs}}, \ and\ \bibinfo
  {author} {\bibfnamefont {G.}~\bibnamefont {Rajagopal}},\ }\href {\doibase
  10.1103/RevModPhys.73.33} {\bibfield  {journal} {\bibinfo  {journal} {Rev.
  Mod. Phys.}\ }\textbf {\bibinfo {volume} {73}},\ \bibinfo {pages} {33}
  (\bibinfo {year} {2001})}\BibitemShut {NoStop}%
\bibitem [{\citenamefont {Nightingale}\ and\ \citenamefont
  {Umrigar}(2014)}]{Nightingale:2014bk}%
  \BibitemOpen
  \bibfield  {author} {\bibinfo {author} {\bibfnamefont {M.}~\bibnamefont
  {Nightingale}}\ and\ \bibinfo {author} {\bibfnamefont {C.~J.}\ \bibnamefont
  {Umrigar}},\ }\href@noop {} {\emph {\bibinfo {title} {Quantum Monte Carlo
  Methods in Physics and Chemistry}}}\ (\bibinfo  {publisher} {Springer},\
  \bibinfo {address} {Vienna},\ \bibinfo {year} {2014})\BibitemShut {NoStop}%
\bibitem [{\citenamefont {Gordillo}\ \emph {et~al.}(2020)\citenamefont
  {Gordillo}, \citenamefont {De~Soto},\ and\ \citenamefont
  {Segovia}}]{Gordillo:2020sgc}%
  \BibitemOpen
  \bibfield  {author} {\bibinfo {author} {\bibfnamefont {M.~C.}\ \bibnamefont
  {Gordillo}}, \bibinfo {author} {\bibfnamefont {F.}~\bibnamefont {De~Soto}}, \
  and\ \bibinfo {author} {\bibfnamefont {J.}~\bibnamefont {Segovia}},\ }\href
  {\doibase 10.1103/PhysRevD.102.114007} {\bibfield  {journal} {\bibinfo
  {journal} {Phys. Rev. D}\ }\textbf {\bibinfo {volume} {102}},\ \bibinfo
  {pages} {114007} (\bibinfo {year} {2020})},\ \Eprint
  {http://arxiv.org/abs/2009.11889} {arXiv:2009.11889 [hep-ph]} \BibitemShut
  {NoStop}%
\bibitem [{\citenamefont {Semay}\ and\ \citenamefont
  {Silvestre-Brac}(1994)}]{Semay:1994ht}%
  \BibitemOpen
  \bibfield  {author} {\bibinfo {author} {\bibfnamefont {C.}~\bibnamefont
  {Semay}}\ and\ \bibinfo {author} {\bibfnamefont {B.}~\bibnamefont
  {Silvestre-Brac}},\ }\href {\doibase 10.1007/BF01413104} {\bibfield
  {journal} {\bibinfo  {journal} {Z. Phys. C}\ }\textbf {\bibinfo {volume}
  {61}},\ \bibinfo {pages} {271} (\bibinfo {year} {1994})}\BibitemShut
  {NoStop}%
\bibitem [{\citenamefont {Silvestre-Brac}(1996)}]{SilvestreBrac:1996bg}%
  \BibitemOpen
  \bibfield  {author} {\bibinfo {author} {\bibfnamefont {B.}~\bibnamefont
  {Silvestre-Brac}},\ }\href {\doibase 10.1007/s006010050028} {\bibfield
  {journal} {\bibinfo  {journal} {Few Body Syst.}\ }\textbf {\bibinfo {volume}
  {20}},\ \bibinfo {pages} {1} (\bibinfo {year} {1996})}\BibitemShut {NoStop}%
\bibitem [{\citenamefont {Fernandez}\ \emph {et~al.}(1993)\citenamefont
  {Fernandez}, \citenamefont {Valcarce}, \citenamefont {Straub},\ and\
  \citenamefont {Faessler}}]{Fernandez:1993hx}%
  \BibitemOpen
  \bibfield  {author} {\bibinfo {author} {\bibfnamefont {F.}~\bibnamefont
  {Fernandez}}, \bibinfo {author} {\bibfnamefont {A.}~\bibnamefont {Valcarce}},
  \bibinfo {author} {\bibfnamefont {U.}~\bibnamefont {Straub}}, \ and\ \bibinfo
  {author} {\bibfnamefont {A.}~\bibnamefont {Faessler}},\ }\href {\doibase
  10.1088/0954-3899/19/12/007} {\bibfield  {journal} {\bibinfo  {journal} {J.
  Phys. G}\ }\textbf {\bibinfo {volume} {19}},\ \bibinfo {pages} {2013}
  (\bibinfo {year} {1993})}\BibitemShut {NoStop}%
\bibitem [{\citenamefont {Valcarce}\ \emph {et~al.}(1996)\citenamefont
  {Valcarce}, \citenamefont {Fernandez}, \citenamefont {Gonzalez},\ and\
  \citenamefont {Vento}}]{Valcarce:1995dm}%
  \BibitemOpen
  \bibfield  {author} {\bibinfo {author} {\bibfnamefont {A.}~\bibnamefont
  {Valcarce}}, \bibinfo {author} {\bibfnamefont {F.}~\bibnamefont {Fernandez}},
  \bibinfo {author} {\bibfnamefont {P.}~\bibnamefont {Gonzalez}}, \ and\
  \bibinfo {author} {\bibfnamefont {V.}~\bibnamefont {Vento}},\ }\href
  {\doibase 10.1016/0370-2693(95)01413-6} {\bibfield  {journal} {\bibinfo
  {journal} {Phys. Lett. B}\ }\textbf {\bibinfo {volume} {367}},\ \bibinfo
  {pages} {35} (\bibinfo {year} {1996})},\ \Eprint
  {http://arxiv.org/abs/nucl-th/9509009} {arXiv:nucl-th/9509009} \BibitemShut
  {NoStop}%
\bibitem [{\citenamefont {Vijande}\ \emph {et~al.}(2005)\citenamefont
  {Vijande}, \citenamefont {Fernandez},\ and\ \citenamefont
  {Valcarce}}]{Vijande:2004he}%
  \BibitemOpen
  \bibfield  {author} {\bibinfo {author} {\bibfnamefont {J.}~\bibnamefont
  {Vijande}}, \bibinfo {author} {\bibfnamefont {F.}~\bibnamefont {Fernandez}},
  \ and\ \bibinfo {author} {\bibfnamefont {A.}~\bibnamefont {Valcarce}},\
  }\href {\doibase 10.1088/0954-3899/31/5/017} {\bibfield  {journal} {\bibinfo
  {journal} {J. Phys. G}\ }\textbf {\bibinfo {volume} {31}},\ \bibinfo {pages}
  {481} (\bibinfo {year} {2005})},\ \Eprint
  {http://arxiv.org/abs/hep-ph/0411299} {arXiv:hep-ph/0411299} \BibitemShut
  {NoStop}%
\bibitem [{\citenamefont {Segovia}\ \emph {et~al.}(2008)\citenamefont
  {Segovia}, \citenamefont {Entem},\ and\ \citenamefont
  {Fernandez}}]{Segovia:2008zza}%
  \BibitemOpen
  \bibfield  {author} {\bibinfo {author} {\bibfnamefont {J.}~\bibnamefont
  {Segovia}}, \bibinfo {author} {\bibfnamefont {D.~R.}\ \bibnamefont {Entem}},
  \ and\ \bibinfo {author} {\bibfnamefont {F.}~\bibnamefont {Fernandez}},\
  }\href {\doibase 10.1016/j.physletb.2008.02.051} {\bibfield  {journal}
  {\bibinfo  {journal} {Phys. Lett. B}\ }\textbf {\bibinfo {volume} {662}},\
  \bibinfo {pages} {33} (\bibinfo {year} {2008})}\BibitemShut {NoStop}%
\bibitem [{\citenamefont {Segovia}\ \emph
  {et~al.}(2011{\natexlab{a}})\citenamefont {Segovia}, \citenamefont
  {Albertus}, \citenamefont {Entem}, \citenamefont {Fernandez}, \citenamefont
  {Hernandez},\ and\ \citenamefont {Perez-Garcia}}]{Segovia:2011dg}%
  \BibitemOpen
  \bibfield  {author} {\bibinfo {author} {\bibfnamefont {J.}~\bibnamefont
  {Segovia}}, \bibinfo {author} {\bibfnamefont {C.}~\bibnamefont {Albertus}},
  \bibinfo {author} {\bibfnamefont {D.~R.}\ \bibnamefont {Entem}}, \bibinfo
  {author} {\bibfnamefont {F.}~\bibnamefont {Fernandez}}, \bibinfo {author}
  {\bibfnamefont {E.}~\bibnamefont {Hernandez}}, \ and\ \bibinfo {author}
  {\bibfnamefont {M.~A.}\ \bibnamefont {Perez-Garcia}},\ }\href {\doibase
  10.1103/PhysRevD.84.094029} {\bibfield  {journal} {\bibinfo  {journal} {Phys.
  Rev. D}\ }\textbf {\bibinfo {volume} {84}},\ \bibinfo {pages} {094029}
  (\bibinfo {year} {2011}{\natexlab{a}})},\ \Eprint
  {http://arxiv.org/abs/1107.4248} {arXiv:1107.4248 [hep-ph]} \BibitemShut
  {NoStop}%
\bibitem [{\citenamefont {Segovia}\ \emph {et~al.}(2009)\citenamefont
  {Segovia}, \citenamefont {Yasser}, \citenamefont {Entem},\ and\ \citenamefont
  {Fernandez}}]{Segovia:2009zz}%
  \BibitemOpen
  \bibfield  {author} {\bibinfo {author} {\bibfnamefont {J.}~\bibnamefont
  {Segovia}}, \bibinfo {author} {\bibfnamefont {A.~M.}\ \bibnamefont {Yasser}},
  \bibinfo {author} {\bibfnamefont {D.~R.}\ \bibnamefont {Entem}}, \ and\
  \bibinfo {author} {\bibfnamefont {F.}~\bibnamefont {Fernandez}},\ }\href
  {\doibase 10.1103/PhysRevD.80.054017} {\bibfield  {journal} {\bibinfo
  {journal} {Phys. Rev. D}\ }\textbf {\bibinfo {volume} {80}},\ \bibinfo
  {pages} {054017} (\bibinfo {year} {2009})}\BibitemShut {NoStop}%
\bibitem [{\citenamefont {Segovia}\ \emph
  {et~al.}(2011{\natexlab{b}})\citenamefont {Segovia}, \citenamefont {Entem},\
  and\ \citenamefont {Fernandez}}]{Segovia:2011zza}%
  \BibitemOpen
  \bibfield  {author} {\bibinfo {author} {\bibfnamefont {J.}~\bibnamefont
  {Segovia}}, \bibinfo {author} {\bibfnamefont {D.~R.}\ \bibnamefont {Entem}},
  \ and\ \bibinfo {author} {\bibfnamefont {F.}~\bibnamefont {Fernandez}},\
  }\href {\doibase 10.1103/PhysRevD.83.114018} {\bibfield  {journal} {\bibinfo
  {journal} {Phys. Rev. D}\ }\textbf {\bibinfo {volume} {83}},\ \bibinfo
  {pages} {114018} (\bibinfo {year} {2011}{\natexlab{b}})}\BibitemShut
  {NoStop}%
\bibitem [{\citenamefont {Segovia}\ \emph {et~al.}(2015)\citenamefont
  {Segovia}, \citenamefont {Entem},\ and\ \citenamefont
  {Fernandez}}]{Segovia:2015dia}%
  \BibitemOpen
  \bibfield  {author} {\bibinfo {author} {\bibfnamefont {J.}~\bibnamefont
  {Segovia}}, \bibinfo {author} {\bibfnamefont {D.~R.}\ \bibnamefont {Entem}},
  \ and\ \bibinfo {author} {\bibfnamefont {F.}~\bibnamefont {Fernandez}},\
  }\href {\doibase 10.1103/PhysRevD.91.094020} {\bibfield  {journal} {\bibinfo
  {journal} {Phys. Rev. D}\ }\textbf {\bibinfo {volume} {91}},\ \bibinfo
  {pages} {094020} (\bibinfo {year} {2015})},\ \Eprint
  {http://arxiv.org/abs/1502.03827} {arXiv:1502.03827 [hep-ph]} \BibitemShut
  {NoStop}%
\bibitem [{\citenamefont {Ortega}\ \emph {et~al.}(2017)\citenamefont {Ortega},
  \citenamefont {Segovia}, \citenamefont {Entem},\ and\ \citenamefont
  {Fern\'andez}}]{Ortega:2016pgg}%
  \BibitemOpen
  \bibfield  {author} {\bibinfo {author} {\bibfnamefont {P.~G.}\ \bibnamefont
  {Ortega}}, \bibinfo {author} {\bibfnamefont {J.}~\bibnamefont {Segovia}},
  \bibinfo {author} {\bibfnamefont {D.~R.}\ \bibnamefont {Entem}}, \ and\
  \bibinfo {author} {\bibfnamefont {F.}~\bibnamefont {Fern\'andez}},\ }\href
  {\doibase 10.1103/PhysRevD.95.034010} {\bibfield  {journal} {\bibinfo
  {journal} {Phys. Rev. D}\ }\textbf {\bibinfo {volume} {95}},\ \bibinfo
  {pages} {034010} (\bibinfo {year} {2017})},\ \Eprint
  {http://arxiv.org/abs/1612.04826} {arXiv:1612.04826 [hep-ph]} \BibitemShut
  {NoStop}%
\bibitem [{\citenamefont {Ortega}\ \emph {et~al.}(2019)\citenamefont {Ortega},
  \citenamefont {Segovia}, \citenamefont {Entem},\ and\ \citenamefont
  {Fern\'andez}}]{Ortega:2018cnm}%
  \BibitemOpen
  \bibfield  {author} {\bibinfo {author} {\bibfnamefont {P.~G.}\ \bibnamefont
  {Ortega}}, \bibinfo {author} {\bibfnamefont {J.}~\bibnamefont {Segovia}},
  \bibinfo {author} {\bibfnamefont {D.~R.}\ \bibnamefont {Entem}}, \ and\
  \bibinfo {author} {\bibfnamefont {F.}~\bibnamefont {Fern\'andez}},\ }\href
  {\doibase 10.1140/epjc/s10052-019-6552-7} {\bibfield  {journal} {\bibinfo
  {journal} {Eur. Phys. J. C}\ }\textbf {\bibinfo {volume} {79}},\ \bibinfo
  {pages} {78} (\bibinfo {year} {2019})},\ \Eprint
  {http://arxiv.org/abs/1808.00914} {arXiv:1808.00914 [hep-ph]} \BibitemShut
  {NoStop}%
\bibitem [{\citenamefont {Ortega}\ \emph {et~al.}(2021)\citenamefont {Ortega},
  \citenamefont {Segovia},\ and\ \citenamefont {Fernandez}}]{Ortega:2021xst}%
  \BibitemOpen
  \bibfield  {author} {\bibinfo {author} {\bibfnamefont {P.~G.}\ \bibnamefont
  {Ortega}}, \bibinfo {author} {\bibfnamefont {J.}~\bibnamefont {Segovia}}, \
  and\ \bibinfo {author} {\bibfnamefont {F.}~\bibnamefont {Fernandez}},\ }\href
  {\doibase 10.1103/PhysRevD.104.094004} {\bibfield  {journal} {\bibinfo
  {journal} {Phys. Rev. D}\ }\textbf {\bibinfo {volume} {104}},\ \bibinfo
  {pages} {094004} (\bibinfo {year} {2021})},\ \Eprint
  {http://arxiv.org/abs/2107.02544} {arXiv:2107.02544 [hep-ph]} \BibitemShut
  {NoStop}%
\bibitem [{\citenamefont {Yang}\ \emph {et~al.}(2019)\citenamefont {Yang},
  \citenamefont {Ping},\ and\ \citenamefont {Segovia}}]{Yang:2018oqd}%
  \BibitemOpen
  \bibfield  {author} {\bibinfo {author} {\bibfnamefont {G.}~\bibnamefont
  {Yang}}, \bibinfo {author} {\bibfnamefont {J.}~\bibnamefont {Ping}}, \ and\
  \bibinfo {author} {\bibfnamefont {J.}~\bibnamefont {Segovia}},\ }\href
  {\doibase 10.1103/PhysRevD.99.014035} {\bibfield  {journal} {\bibinfo
  {journal} {Phys. Rev. D}\ }\textbf {\bibinfo {volume} {99}},\ \bibinfo
  {pages} {014035} (\bibinfo {year} {2019})},\ \Eprint
  {http://arxiv.org/abs/1809.06193} {arXiv:1809.06193 [hep-ph]} \BibitemShut
  {NoStop}%
\bibitem [{\citenamefont {Yang}\ \emph
  {et~al.}(2020{\natexlab{a}})\citenamefont {Yang}, \citenamefont {Ping},\ and\
  \citenamefont {Segovia}}]{Yang:2020twg}%
  \BibitemOpen
  \bibfield  {author} {\bibinfo {author} {\bibfnamefont {G.}~\bibnamefont
  {Yang}}, \bibinfo {author} {\bibfnamefont {J.}~\bibnamefont {Ping}}, \ and\
  \bibinfo {author} {\bibfnamefont {J.}~\bibnamefont {Segovia}},\ }\href
  {\doibase 10.1103/PhysRevD.101.074030} {\bibfield  {journal} {\bibinfo
  {journal} {Phys. Rev. D}\ }\textbf {\bibinfo {volume} {101}},\ \bibinfo
  {pages} {074030} (\bibinfo {year} {2020}{\natexlab{a}})},\ \Eprint
  {http://arxiv.org/abs/2003.05253} {arXiv:2003.05253 [hep-ph]} \BibitemShut
  {NoStop}%
\bibitem [{\citenamefont {Yang}\ \emph
  {et~al.}(2020{\natexlab{b}})\citenamefont {Yang}, \citenamefont {Ping},\ and\
  \citenamefont {Segovia}}]{Yang:2020fou}%
  \BibitemOpen
  \bibfield  {author} {\bibinfo {author} {\bibfnamefont {G.}~\bibnamefont
  {Yang}}, \bibinfo {author} {\bibfnamefont {J.}~\bibnamefont {Ping}}, \ and\
  \bibinfo {author} {\bibfnamefont {J.}~\bibnamefont {Segovia}},\ }\href
  {\doibase 10.1103/PhysRevD.102.054023} {\bibfield  {journal} {\bibinfo
  {journal} {Phys. Rev. D}\ }\textbf {\bibinfo {volume} {102}},\ \bibinfo
  {pages} {054023} (\bibinfo {year} {2020}{\natexlab{b}})},\ \Eprint
  {http://arxiv.org/abs/2007.05190} {arXiv:2007.05190 [hep-ph]} \BibitemShut
  {NoStop}%
\bibitem [{\citenamefont {Meyer}\ and\ \citenamefont
  {Swanson}(2015)}]{Meyer:2015eta}%
  \BibitemOpen
  \bibfield  {author} {\bibinfo {author} {\bibfnamefont {C.~A.}\ \bibnamefont
  {Meyer}}\ and\ \bibinfo {author} {\bibfnamefont {E.~S.}\ \bibnamefont
  {Swanson}},\ }\href {\doibase 10.1016/j.ppnp.2015.03.001} {\bibfield
  {journal} {\bibinfo  {journal} {Prog. Part. Nucl. Phys.}\ }\textbf {\bibinfo
  {volume} {82}},\ \bibinfo {pages} {21} (\bibinfo {year} {2015})},\ \Eprint
  {http://arxiv.org/abs/1502.07276} {arXiv:1502.07276 [hep-ph]} \BibitemShut
  {NoStop}%
\end{thebibliography}%

\end{document}